\documentclass{article}
\usepackage{enumitem}
\usepackage{spconf,amsmath,graphicx,amssymb}
\usepackage{color}
\usepackage{footnote}
\makesavenoteenv{tabular}
\makesavenoteenv{table}
\PassOptionsToPackage{hyphens}{url}\usepackage{hyperref}

\hypersetup{colorlinks=true}

\title{Tuplemax Loss for Language Identification}
\name{Li Wan \qquad Prashant Sridhar \qquad Yang Yu \qquad Quan Wang \qquad Ignacio Lopez Moreno}

\address{Google Inc., USA\\[4pt] {
  \normalsize
  \{\href{mailto:liwan@google.com}{\nolinkurl{liwan}},
    \href{mailto:psridhar@google.com}{\nolinkurl{psridhar}},
    \href{mailto:yyuyy@google.com}{\nolinkurl{yyuyy}},
    \href{mailto:quanw@google.com}{\nolinkurl{quanw}},
    \href{mailto:elnota@google.com}{\nolinkurl{elnota}}\}
    {\tt @google.com}
}}

\begin{document}
\ninept
\maketitle
\begin{abstract}
In many scenarios of a language identification task, the user will specify a small set of languages which he/she can speak instead of a large set of all possible languages. 
We want to model such prior knowledge into the way we train our neural networks, by replacing the commonly used softmax loss function with a novel loss function named \emph{tuplemax loss}. As a matter of fact, a typical language identification system launched in North America has about $95\%$ users who could speak no more than two languages. 
Using the tuplemax loss, our system achieved a 
$2.33$\% error rate, which is a relative $39.4$\% improvement over the $3.85\%$ error rate of standard softmax loss method.
\end{abstract}
\section{Introduction}
\label{sec:intro}



Large vocabulary continuous speech recognition (LVCSR) systems are becoming increasingly relevant for industry, tracking the technological trend toward increased human interaction using voice-operated devices~\cite{schalkwyk2010your}. However, even for accurate LVCSR systems, certain obstacles can diminish the user experience. For multilingual speakers, one such obstacle is the monolingual character of LVCSR systems, meaning that users are limited to speaking a single, preset language. One way to solve this problem is to use a spoken Language Identification (LID) system to select the most likely language in the spoken utterance~\cite{gonzalez2015real}. In some cases, multilingual LVCSR systems can also take advantage of the fact that most multilingual speakers only speak a limited number of languages, $n$. For example, the user can be asked to pre-select $n$ candidate languages from the super-set of all $N$ available languages. By conditioning the multilingual LVCSR system to a tuple of $n$ candidate languages, the LID system can condition its decision to it. The LID system could, for example, discard other potentially misleading languages. We refer to this problem as the Tuple-Conditioned Multi-Class Classification (TCMCC) problem. We argue that most LID systems optimized using the softmax cross-entropy loss (we will refer to it as \emph{softmax loss} for simplicity) may be sub-optimal for the TCMCC problem, as the softmax loss only optimizes the evaluation metric for the cases that $n=N$. To solve this problem, we extend the idea of conditioning the LID system to the training stage by introducing a new loss that directly optimizes the TCMCC problem. We refer to this loss as \emph{tuplemax loss}. Moreover, we show that as the tuple size grows $n \rightarrow N$, the tuplemax and the softmax losses converge asymptotically, which complies with a provable property that tuplemax can be regarded as a generalization of the softmax loss.

The proposed tuplemax loss is developed in the context of our LID system, whose architecture presented in this paper is extended from the ideas originally proposed in~\cite{gonzalez2014automatic}~\cite{gonzalez2015frame}~\cite{lopez2016use}.
Most modern LID systems are composed of two-stage processing. First, a fixed-length vector representation of the utterance (\emph{i.e.} embedding) is used to encode the language information of the input sequence. Then, a discriminative $N$-class classifier is trained on the space of the embeddings to generate scores for each language~\cite{fer2015multilingual}~\cite{martinez2011language}~\cite{snyder2018spoken}. So far the most successful types of embeddings are those that are computed using neural networks~\cite{snyder2018spoken}, which have shown to be able to match or surpass the performance of the more traditional i-vector embeddings~\cite{snyder2018spoken}, particularly in short utterances. In our LID model, which is based on a multi-layer deep Long Short-Term Memory (LSTM) architecture, the embedding vector is given by the last activation vector of the sequence. This temporal pooling operation also allows the system to provide scores with minimal latency and without requiring any right context or padding. The second stage is usually governed by a discriminative Gaussian classifier trained on the space of the embeddings, potentially followed by an additional calibrating system~\cite{martinez2011language}. In our model, we choose to directly use the probabilities produced by the last layer~\cite{lopez2014automatic}. This is an approach that provides a simpler end-to-end optimization. However, we believe that the tuplemax loss could also be beneficial for some of the systems based on loosely connected classifiers, particularly in the cases where the network computing the embeddings is optimized using the cross-entropy loss of a softmax layer, like~\cite{snyder2018spoken}.



\section{Tuplemax Loss}
\label{sec:tuplemax}

\begin{figure*}[ht]
  \centering
    \includegraphics[width=0.85\textwidth]{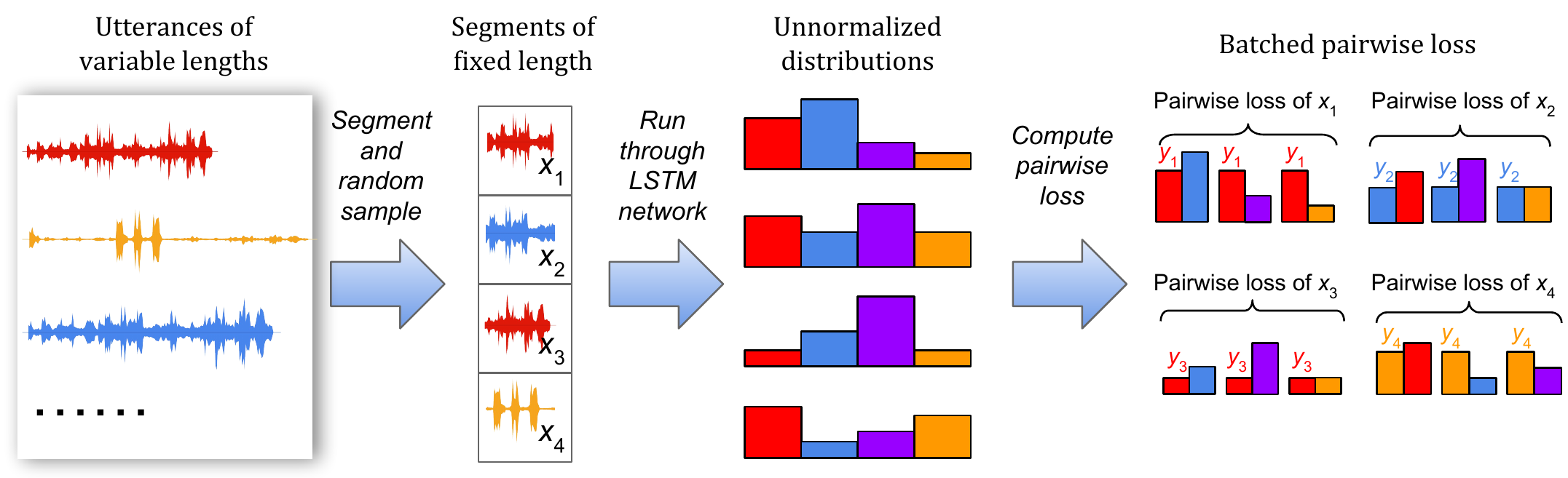}
  \vspace{-1em}
  \caption{
    System overview. Different colors indicate utterances/embeddings of different languages.
  }
  \label{fig:overview}
\end{figure*}
Each training utterance is first 
processed to generate a feature sequence $\mathbf{x}$ with 40-dimensional log-mel-filterbank 
and a label $y$, where $y \in \{1,\ldots, N\}$ and $N$ is the number of target languages.
Let $\mathbf{z}$ be the $N$-dimensional output of the last layer of the network. $\mathbf{z} = f(\mathbf{x};\mathbf{w})$ represents the unnormalized distribution of input $\mathbf{x}$ over the $N$ target languages, where $\mathbf{w}$ represents the parameters of the network. Let $\mathbf{z}_k$ be the predicted unnormalized probability of the $k^{\mathrm{th}}$ language. 

The general language identification problem is formulated as a standard multi-class classification problem, which maximizes the probability of the correct label and minimizes that of all the others, {\it equally}. This is represented as Eq.~(\ref{eqn:softmax}),
where $z$ is unnormalized logits.
\begin{equation}
\label{eqn:softmax}
L(y, \mathbf{z})=-\log \frac{\exp(\mathbf{z}_y)}{\sum_k \exp(\mathbf{z}_k)}=\log\sum_{k=1}^N \exp(\mathbf{z}_k)-\mathbf{z}_{y}.
\end{equation}

However, at inference time, the system must choose from only two or three languages ($\mathbf{S}$) which the user has picked in their settings beforehand. There are exponential number of possible language combinations that the users may choose. It is impractical to train individual models for each possible language set. Therefore, we must train a model that can output probabilities for each of the $N$ supported languages and have the inference system pick the best language from the user selected subset $\mathbf{S}$.
This discrepancy between $N$, the number of supported languages, and $\mathbf{S}$, the user selected subset means that standard softmax loss might not be optimal for our problem. The majority of multilingual speakers speak two languages. This makes the pairwise classification particularly interesting. 

\vspace{0.3cm}
\noindent\framebox{
\begin{minipage}{25em}
{\bf Practical Example} 
Consider two distributions with the first language as ground truth:
$[0.3, 0.4, 0.2, 0.1]$ and $[0.3, 0.25, 0.25, 0.2]$. The softmax cross-entropy loss for both outputs is $-\log 0.3$.
However, the second output is significantly better than the first one, since the system will always return the correct language for all the language pairs that contain the true language and any other one based on the second output. In contrast, the first output will have one incorrect pair.    
\end{minipage}
}
\vspace{0.3cm}

In order to model such pairwise relationship, we have pairwise loss defined in Eq.~(\ref{eqn:loss_pairwise}):
\begin{eqnarray}
L(y,\mathbf{z})&=&-\mathbf{E}_{k\neq y}[\log\frac{\exp(\mathbf{z}_y)}{\exp(\mathbf{z}_y)+\exp(\mathbf{z}_k)}] \nonumber \\
      &=&\mathbf{E}_{k\neq y}[\log(\exp(\mathbf{z}_y)+\exp(\mathbf{z}_k))]-\mathbf{z}_y .
      \label{eqn:loss_pairwise}
\end{eqnarray}
where $\mathbf{E}_s$ denotes the expection of a set $s$. 
A subset $\mathbf{S}_y^n$ is a collection of all the tuples that includes the correct label $y$ with $n$ elements with $1<n\leq N$, the loss becomes:
\begin{equation}
\label{eqn:loss_tuple}
L^n(y,\mathbf{z})={\bf E}_{\mathbf{S}^n_y}[log \sum_{k\in \mathbf{S}^n_y}\exp(\mathbf{z}_k)]-\mathbf{z}_y ,
\end{equation}
where $\mathbf{S}^n_y$ is all the tuples in $\mathbf{S}_y$ with size $n$.
There are $\|S^n\| = \binom{N-1}{n-1}$ tuples in $S^n_y$. In particular, the pairwise loss and the softmax cross-entropy loss of multi-class classification are two special cases:
\begin{enumerate}
    \item $n=2$: we know there are $N-1$ terms in $\|\mathbf{S}^2_y\|$ if we only consider pairs. 
    Eq.~(\ref{eqn:loss_tuple}) reduce to pairwise loss as shown in Eq.~(\ref{eqn:loss_pairwise}). 
    \item $n=N$: there are only $1$ term in $\mathbf{S}_y^N$, Eq.~(\ref{eqn:loss_tuple}) reduce to 
    normal cross-entropy softmax as shown in Eq.~(\ref{eqn:softmax}). In typical
    classification problem (such as image classification), we do not have prior knowledge that
    data is draw from some subset. In this case, softmax cross-entropy is the optimal.
\end{enumerate}
Eq.~(\ref{eqn:loss_tuple}) defines a loss when we want to produce a label from a subset
of $n$ labels. The total loss $L(y,\mathbf{z})$ is a weighted sum of all losses of different sizes for all $1<n\leq N$-language case based on their probabilities, we called
it {\it tuplemax loss}:
\begin{equation}
    \label{eqn:loss_all}
    L(y,\mathbf{z})=\mathbf{E}_{\mathbf{S}^n\sim D}[L^n(y, \mathbf{z})]=\sum_{n=2}^N p_n L^n(y,\mathbf{z}) ,
\end{equation}
\noindent where $p_n$ is the probability of tuples
of size $n$ and $L^n(y, \mathbf{z})$ is loss associate with it.

If we have a prior knowledge
about subset size $\|S^n\|$ distribution $D$, the most optimal
loss function is Eq.~(\ref{eqn:loss_all}). If
such set is always a set of all labels ${1,\ldots,N}$, we should
use softmax loss. 
In practice, more than $95\%$ of the users only specify two languages. 
Currently, we train and evaluate our model focus on pairwise relationship
in our experiments. This is only a special case of tuplemax loss defined in Eq.~(\ref{eqn:loss_all}).

\section{Model Description}
\label{model}
Our LID model is trained to distinguish among 79 different target languages. 
To speed up the training process, we truncate utterances that are longer than 4 seconds. This should cover most of the utterances in full, as we have seen from our query \emph{VoiceSearch} logs ~\cite{schalkwyk2010your}, that the average utterance length is 4.3 seconds. This process will make our input sequences to be 400 samples long, at most. We then compute unnormalized logits over the target languages through a stack of LSTM networks~\cite{hochreiter1997long}.
The loss function is defined over these unnormalized logits for the correct label against all other wrong labels, as is depicted in Figure~\ref{fig:overview}.

\subsection{Network Architecture}
\label{topology}


The first layer of the network is called a \emph{concatenation layer}, which concatenates every two neighboring frames, thus doubling the frame dimension and halving the sequence length.
With this concatenation layer, although we result in more parameters in the bottom LSTM layer, we can significantly speed up training since we only have to unroll the LSTM 200 steps instead of 400 steps.

Following the concatenation layer is a stack of 4 LSTM layers.
Each LSTM layer, except the last one, is followed by a projection layer of 256 units. The number of cells in each LSTM layer can be found in Table~\ref{tab:nn}. For example, ``LSTM(1024, 256)'' denotes an LSTM layer with 1024 memory cells and a projection layer with 256 cells. Our LSTM layers also have a pyramid-alike shape, where the bottom layer has 1024 memory cells and the top layer has only 256 memory cells. Experimentally, we find that both adding projection layers and reducing the size of layers further up in the network significantly speed up training and inference without hurting performance. After the LSTM layers is a temporal pooling layer that simply takes the last activation of the LSTM outputs, followed by ReLU and a linear projection to the number of languages.

\begin{table}[t]
  \caption{The LSTM network architecture.}
  \label{tab:nn}
  \begin{center}
  \scalebox{0.9}{
  \begin{tabular}{ l|l|l}
    \hline
    Index & Input $\rightarrow$ Output Size & Layer Specification \\
    \hline
    0 & $40\times 400\rightarrow 80\times 200$ & Concatenation \& subsampling \\
    1 & $80\times 200\rightarrow 256\times 200$ & LSTM(1024, 256) \\
    2 & $256\times 200\rightarrow 256\times 200$ & LSTM(768, 256) \\
    3 & $256\times 200\rightarrow 256\times 200$ & LSTM(512, 256) \\
    4 & $256\times 200\rightarrow 256\times 200$ & LSTM(256, ---) \\
    5 & $256\times 200\rightarrow 256$ & Last frame outpout \\
    6 & $256\rightarrow 256$ & ReLU  \\
    7 & $256\rightarrow 79$ & Linear \\
    \hline
  \end{tabular}
  }
  \end{center}
  \vspace{-3em}
\end{table}

\subsection{Inference}
At inference time, the user has specified a subset $\mathbf{S}\subseteq \{1,\ldots,N\}$ of languages over which our decision must be made within. Our prediction of the spoken language $y^*$ can be represented as:
$y^*=\arg\max_{k\in \mathbf{S}} f(\mathbf{x};\mathbf{w}) = \arg\max_{k\in \mathbf{S}} \mathbf{z}_k .$

\begin{figure}[ht]
  \centering
    \includegraphics[width=0.4\textwidth]{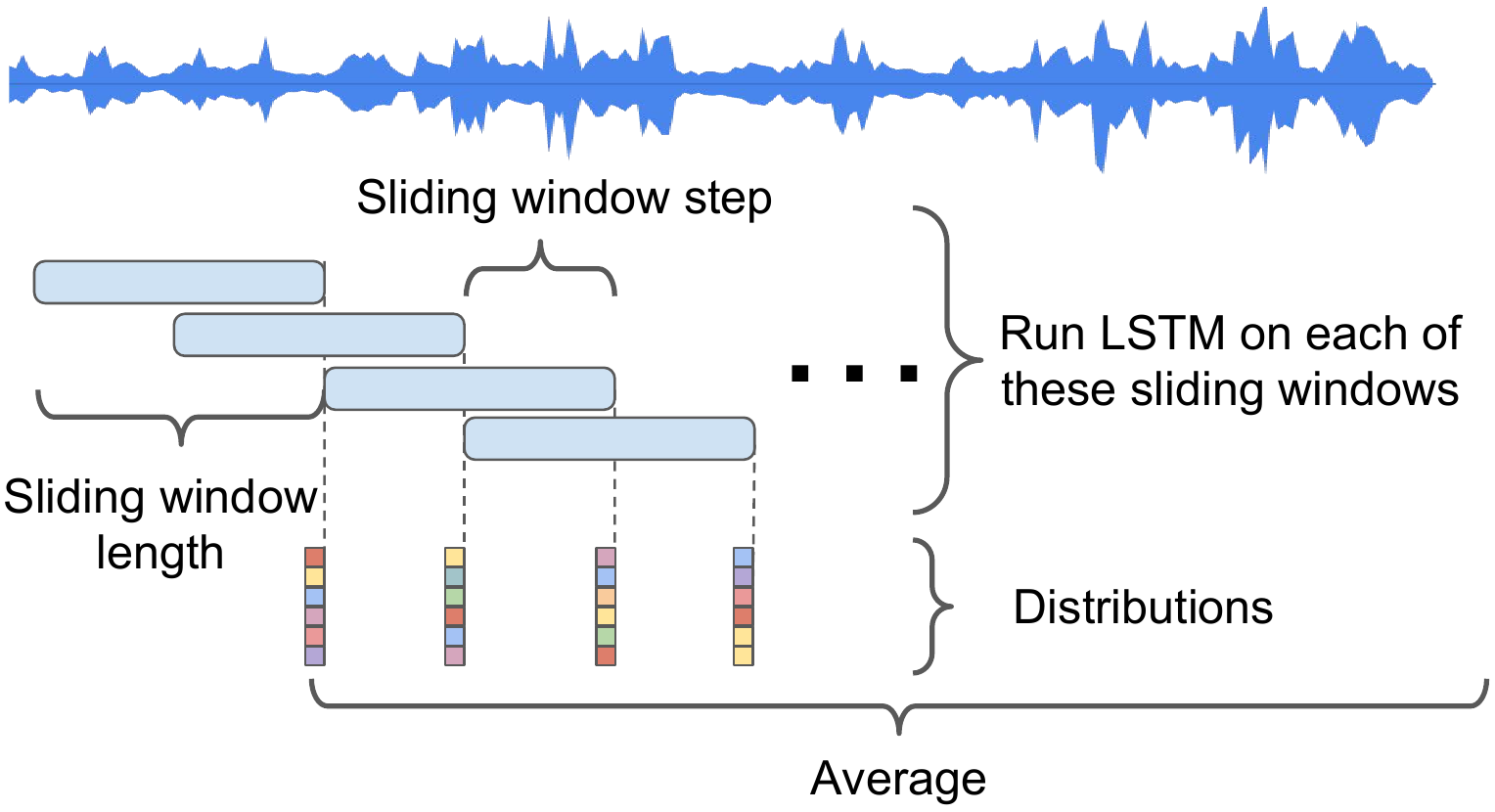}
  \vspace{-1em}
  \caption{
      We aggregate segment inference results for a long utterance.
  }
  \label{fig:inference}
  \vspace{-1em}
\end{figure}

At inference time, the utterance length can vary. Following the standard practice in speaker recognition~\cite{Wan2018GeneralizedEL}~\cite{rahman2018attention}~\cite{wang2018speaker}~\cite{zhang2018fully},
we divide long utterances into overlapping windows of fixed length. The final output is an average of the network response at the end of each fixed length window (shown in Figure~\ref{fig:inference}):
\begin{equation}
\label{eqn:inference_full}
y^*=\arg\max_{k\in \mathbf{S}}\mathbf{E}_{t} [f(\mathbf{x}^t;w)]=\arg\max_{k\in \mathbf{S}}\mathbf{E}_{t} [\mathbf{z}_k^t] ,
\end{equation}
where $\mathbf{x}^t$ is the input segment for the $t^\mathrm{th}$ sliding window and $\mathbf{z}^t$ is the corresponding network response.
This sliding windowing approach allows the system to be more robust to long utterances where the state vector of LSTM network may diverge if the LSTM is unrolled in full.

\section{Experiments}
\label{sec:exp}

\subsection{Training and Evaluation Setup}
\label{data}
Our model is trained to distinguish between $79$ different languages.
Each language has a training set varying from 1M to 60M utterances and a evaluation set of 20K utterances. 
All evaluation utterances and part of the training utterances are from a supervised set transcribed by human transcribers.
Other training utterances are collected from anonymized user voice queries, where the language label is obtained from either user setting or a previous LID system. We remove the non-speech parts from and add procedurally generated noise to the training utterances before training. Both training and evaluation data are mostly collected from monolingual speakers.


On the evaluation data, we compare the performance of the traditional softmax loss and the proposed tuplemax loss, where in the tuplemax loss we use a tuple size of two (pairwise loss, $n=2$), since most of our multilingual traffic has two languages.

For the evaluation metrics, we care about the averaged accuracy/error for \emph{pairwise classification}: we compute the accuracy/error for many two-language binary classification tasks, and report the average value on all the two-language pairs. Specifically, we report both the averaged errors on \textit{all languages pairs} ($79\times 79$ pairs in total), and on $87$ most frequent \textit{top language pairs}. 
For example, ``es-US vs. en-US'' is one of the top language pairs, meaning a significant amount of our users in US speak both Spanish and English.



\subsection{Pairwise Confusion Matrices}
\label{sec:confusion}
\begin{figure*}[ht]
  \centering
    \includegraphics[width=0.85\textwidth]{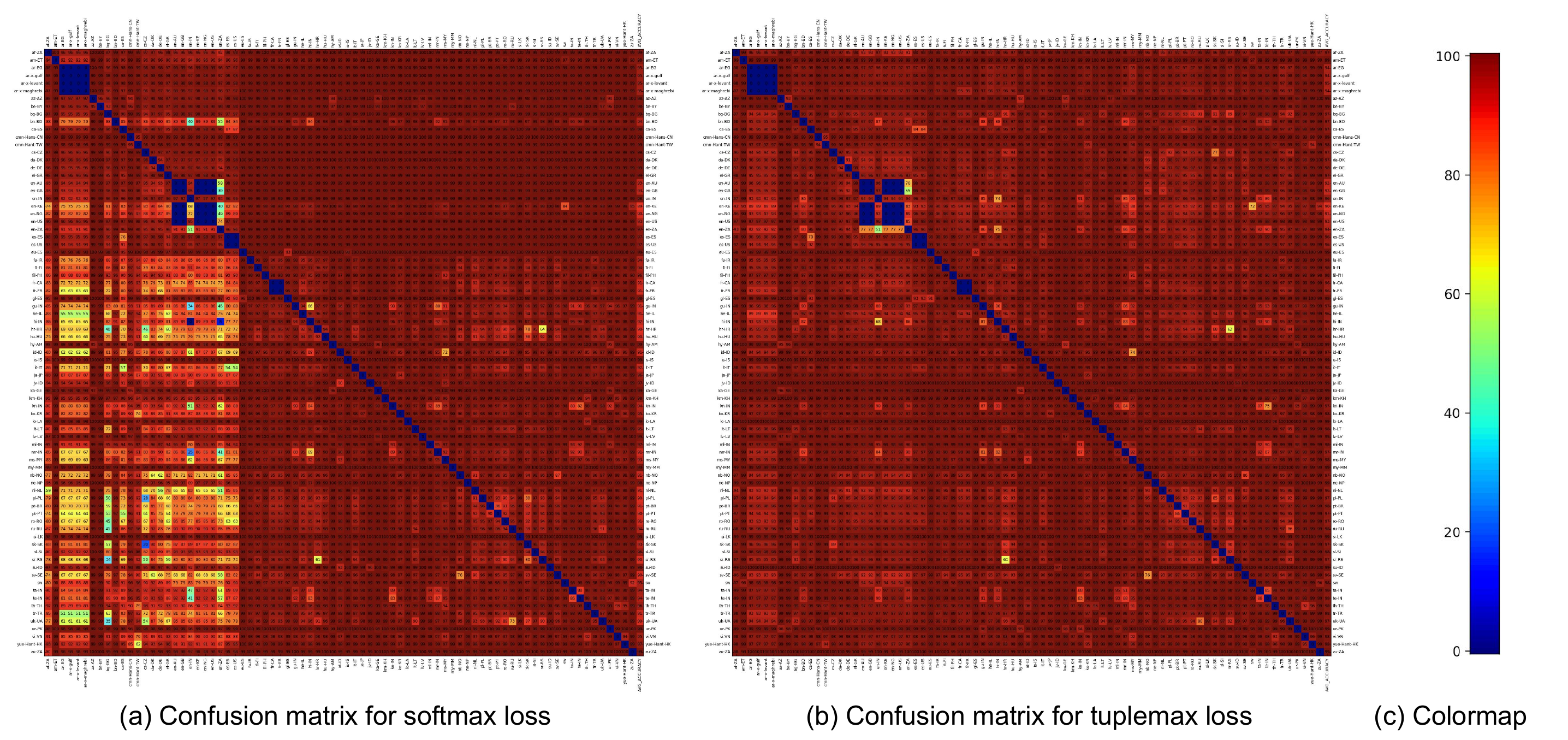}
    \vspace{-0.5em}
  \caption{
    Pairwise confusion matrices for different loss functions on the evaluation data (a and b) and colormap associated with the pairwise confusion matrix accuracy (c). The main purpose of this figure is to illustrate how the tuplemax loss can better balance the classification result than the softmax loss. 
    For example, ``en-AU'', ``en-GB'', ``en-KE'', ``en-US'' are all treated as ``English'', resulting in a larger blue square for ``en'' than other languages. The reader is encouraged to zoom in this high resolution image to better read the labels of each pair.
  }
  \label{fig:evaluation_matrix}
  \vspace{-0.5em}
\end{figure*}

In Fig.~\ref{fig:evaluation_matrix}~(a) and~(b), we present the language classification confusion matrices of the softmax model and the tuplemax model, respectively.
Each element $E_{ji}$ in the confusion matrix represent the classification accuracy in a subset where the utterances have ground truth label $j$, and the user has selected both language $j$ and language $i$ as his/her preferred languages.
For example, if $E_{ji}=90$, it means that $90\%$ of the utterances with ground truth label $j$ and user preference $\mathbf{S}=\{j,i\}$ are correctly classified as language $j$.
The diagonal elements always equal to $0$
because $E_{jj}$ is meaningless. 

%

In Fig.~\ref{fig:evaluation_matrix}~(a), we can see there are several big ``yellow'' clusters on the left half of the image. This implies that
softmax loss, despite being optimal for the N-class classification, leads to {\it unbalanced classification} results when $n=2$. This is because pairwise accuracy is not being modeled by softmax loss function, which is similar to the practical example given in Section~\ref{sec:tuplemax}.
We also notice that those ``yellow'' clusters do not exist in Fig.~\ref{fig:evaluation_matrix}~(b), which indicates that the
classification accuracy is well {\it balanced} for tuplemax loss.




\subsection{Evaluation Results}
\label{sec:error_during_training}
\begin{figure}[ht]
  \centering
    \includegraphics[width=0.43\textwidth]{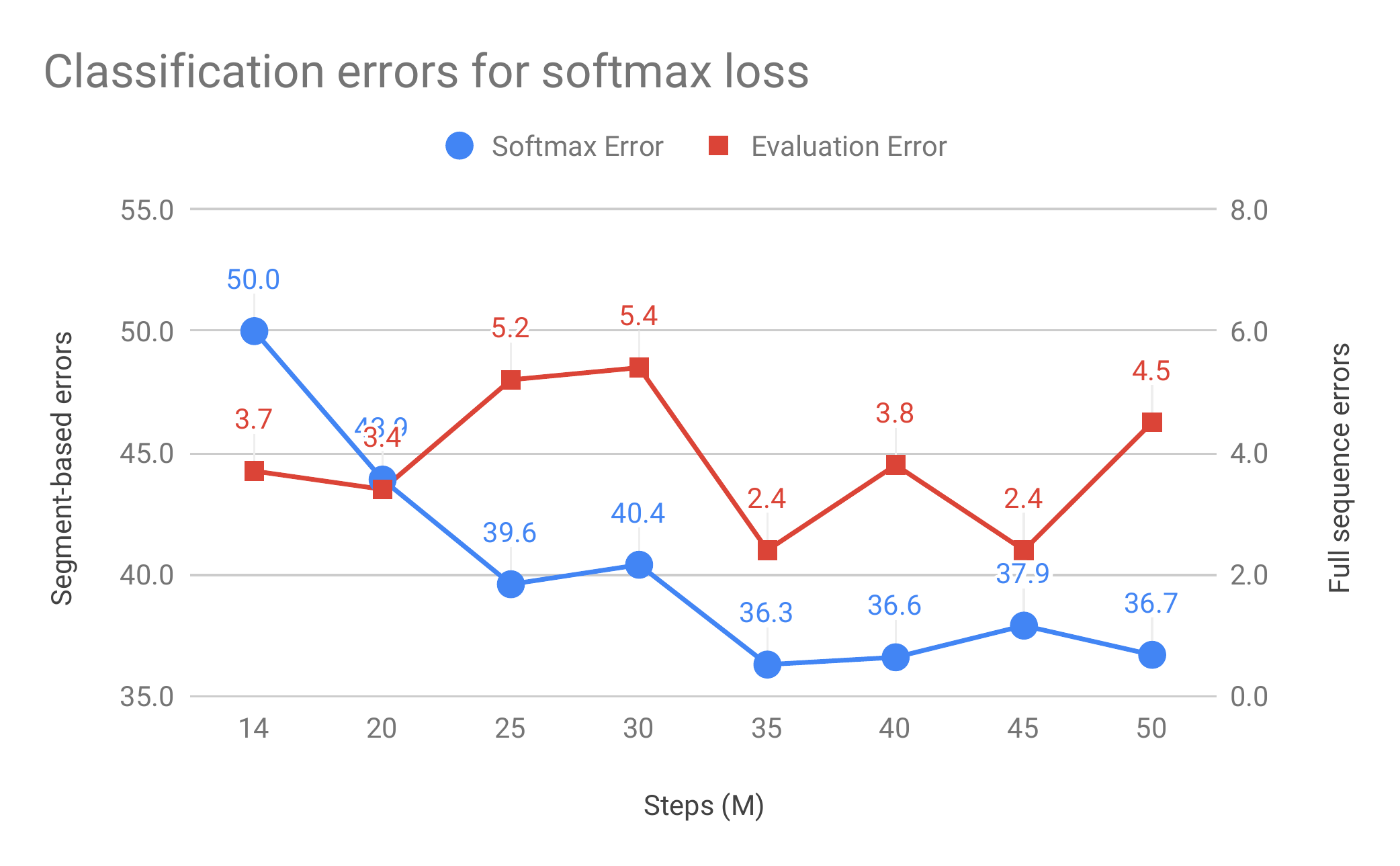}
  \caption{Softmax model convergence at different training steps. Convergence is shown by the relative reduction of the the top-one N-class classification error rate when $N=79$.}
  \label{fig:softmax_eval}
\end{figure}

\begin{figure}[ht]
  \centering
    \includegraphics[width=0.43\textwidth]{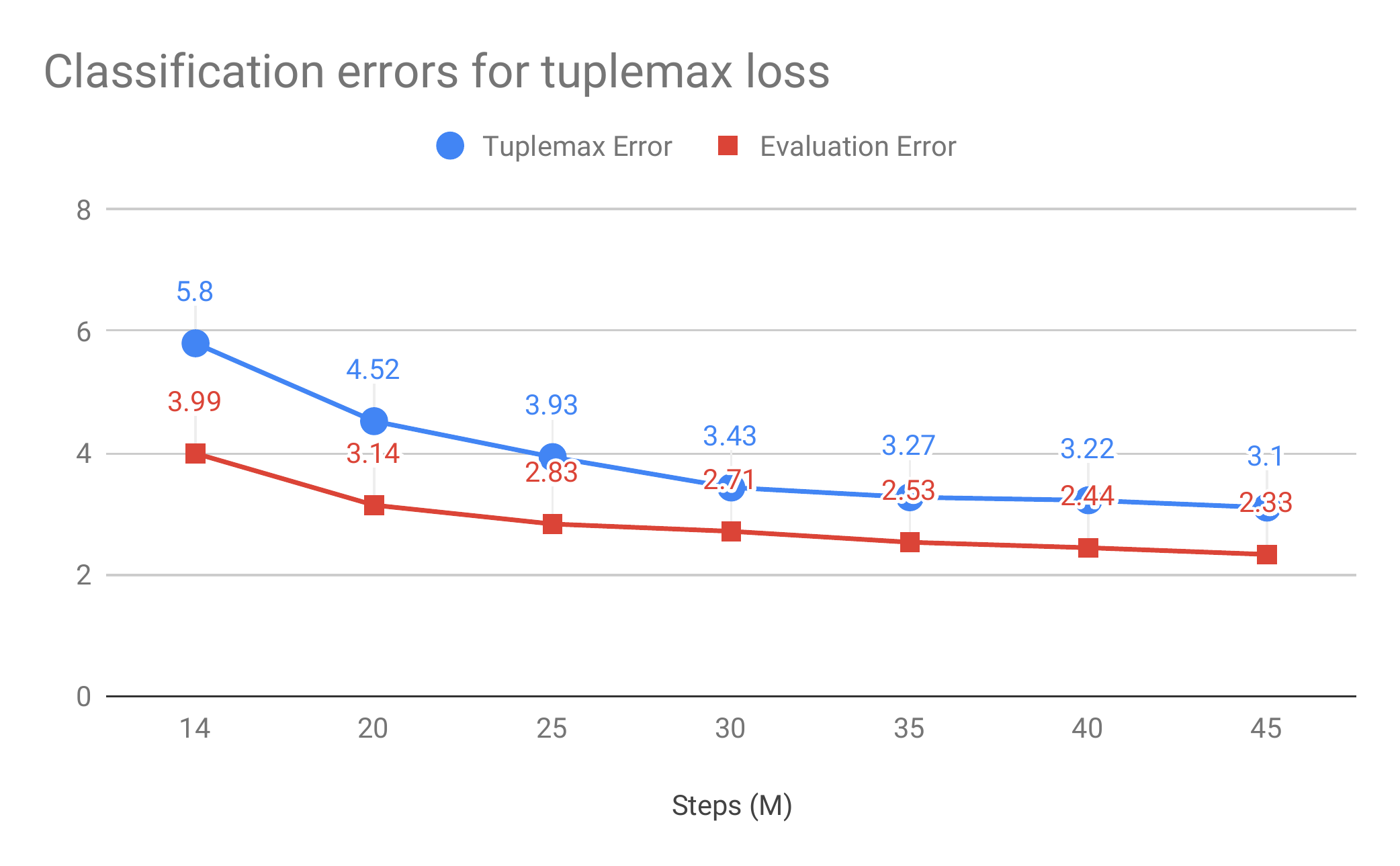}
  \caption{Tuplemax model convergence at different training steps. Both tuplemax error and evaluation error decreases as we train network longer.}
  \label{fig:tuplemax_eval}
  \vspace{-1em}
\end{figure}

Fig.~\ref{fig:softmax_eval} shows that when the softmax model is exposed to the standard top-one N-class classification task ($n=N=79$) can effectively produce steady improvements over time. However there is a lack of correlation top-1 classification error and evaluation error. From Fig~\ref{fig:tuplemax_eval}, we can see tuplemax error is consistent with evaluation error.

\begin{figure}[ht]
  \centering
    \includegraphics[width=0.43\textwidth]{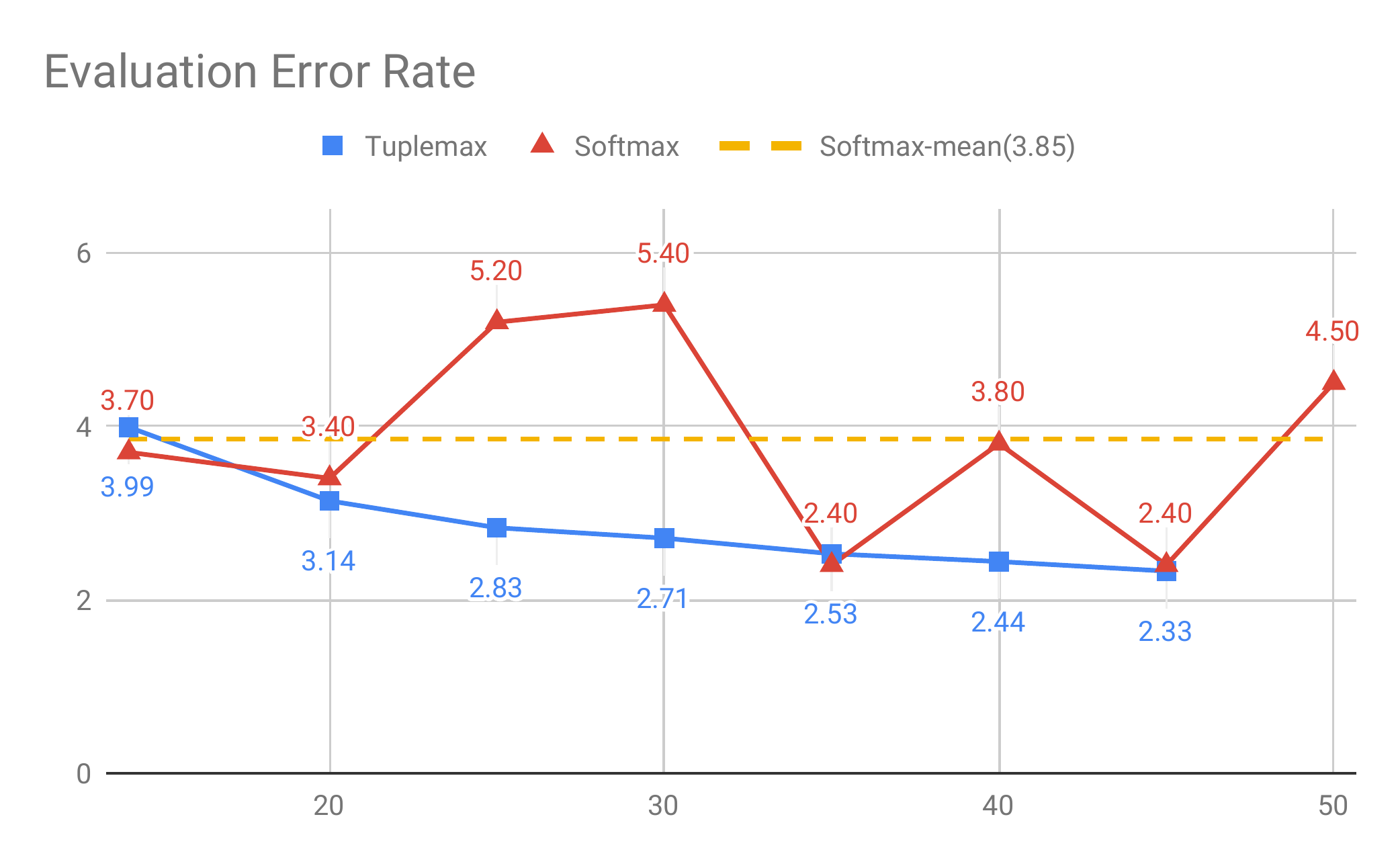}
    \vspace{-0.5em}
  \caption{Pairwise classification errors for tuplemax loss (blue) and softmax loss (red) at different number of training steps. The mean error rate of the softmax loss is also shown in yellow. The tuplemax loss is able to provide steady improvements.}
  \label{fig:checkpoint_eval}
  \vspace{-1em}
\end{figure}

In Fig.~\ref{fig:checkpoint_eval}, we compare the performance of the tuplemax and softmax losses at different number of training steps. The network architecture is kept fixed for both models, softmax and tuplemax models, while the learning rate is optimized  o achieve the best possible training accuracy in both cases. Results show that tuplemax loss can effectively provide steady improvements over time, reaching a value of $2.33\%$. In contrast, the results of the softmax loss oscillate, upon convergence, around a mean error rate of $3.85\%$. This oscillation of the pairwise error rate of the softmax model can make the process of training LID models difficult, and require constant evaluations during training. In our evaluation results, we saw that the error rate of the softmax model can increase by even a factor of 2 in checkpoints that close in time. 

In Table~\ref{tab:error_compare} we also compare the tuplemax model performance with the softmax model performance using multiple strategies for model checkpoint selection. At its best, the softmax model performance still falls behind the performance of the tuplemax model for both, \emph{all} and \emph{top} language pairs. 

\section{Conclusions}
\label{conclusion}
In this paper, we propose a novel loss function named \emph{tuplemax loss}, which is designed for the Tuple-Conditioned Multi-Class Classification problem, a type of classification task where the decision is restricted to a known subset of candidates during inference time. 
We've shown in experiments that tuplemax loss is preferred over softmax loss as $a)$ Tuplemax produces better and balanced results and $b)$ convergence is significantly more stable.

\begin{table}[ht]
  \caption{
    Classification Error Rate (\%) for softmax and tuplemax.
  }
  \label{tab:error_compare}
  \begin{center}
  \begin{small}
  \begin{tabular}{l|l|l|l}
    \hline
    Loss  & Checkpoint & all language & top language \\
    function & type & pairs & pairs  \\
    \hline
     & Last & 4.50 & 11.1 \\
    Softmax & Average & 3.85 & 9.14 \\
     & Best on {\it test}
     \footnote{Picking the best checkpoint based on test data is cheating and it does not reflect the actual model performance.
     Averaging all checkpoints is more fair to estimate model performance.
     Reporting the best checkpoint on test data for softmax is only to illustrate tuplemax is better than all checkpoints produced by softmax.} & 2.40 & 5.50 \\
    \hline
    Tuplemax & Last & {\bf 2.33} & {\bf 4.55} \\
    \hline
  \end{tabular}
  \end{small}
  \end{center}
  \vspace{-1em}
\end{table}

\clearpage
\newpage
\bibliographystyle{IEEEbib}
\bibliography{refs}
\end{document}